\documentclass{appolb}
\usepackage{graphicx}
\usepackage{cite}

\begin{document}
\eqsec  
\title{OFF-SHELL $t\bar{t}j$ PRODUCTION AND TOP QUARK MASS STUDIES AT THE LHC
\thanks{Presented by G.~Bevilacqua at the XLI International Conference of Theoretical Physics "Matter To The Deepest", Podlesice, Poland, September 3-8, 2017}
\thanks{Preprint number: TTK-17-38}
} 
\author{\vspace{0.2cm} Giuseppe Bevilacqua
\address{MTA-DE Particle Physics Research Group, University of Debrecen, H-4010 Debrecen, PBox 105, Hungary}
\\ \vspace{0.5cm}
{Heribertus Bayu Hartanto
\address{Institute for Particle Physics Phenomenology, Department of Physics, Durham University, Durham, DH1 3LE, UK}
}
\\ \vspace{0.5cm}
{Manfred Kraus, Markus Schulze
\address{Humboldt-Universit\"at zu Berlin, Institut f\"ur Physik, Newtonstra\ss{}e 15, D-12489 Berlin, Germany}
}
\\ \vspace{0.5cm}
{Malgorzata Worek
\address{Institute for Theoretical Particle Physics and Cosmology, RWTH Aachen University, D-52056 Aachen, Germany}
}
}

\maketitle
\begin{abstract}
Precision studies of the properties of the top quark represent a cornerstone of the LHC physics program. In this contribution we focus on the production of $t\bar{t}$ pairs in association with one hard jet and in particular on its connection with precision measurements of the top quark mass at the LHC. We report a summary of a full calculation of the process $pp \to e^+\nu_e\mu^-\bar{\nu}_\mu b \bar{b}j$ at NLO QCD accuracy, which describes $t\bar{t}j$ production with leptonic decays beyond the Narrow Width Approximation (NWA), and discuss the impact of the off-shell effects through comparisons with NWA. Finally we explore the sensitivity of $t\bar{t}j$ in the context of top-quark mass extraction with the template method, considering two benchmark observables as case studies.
\end{abstract}

\section{Introduction}
With its unprecedented values of luminosity and center-of-mass energy, the Large Hadron Collider (LHC) has all the features of a top factory: $t\bar{t}$ events are produced abundantly allowing to study the properties of top quarks with high precision. 
The cross section of the inclusive process $pp \to t\bar{t}+X$ is an important benchmark of the Standard Model (SM) with a wealth of phenomenological applications. Precision tests of perturbative QCD \cite{Czakon:2013goa}, constraints on large-$x$ parton distribution functions (PDF) \cite{Czakon:2016olj} and accurate determinations of SM parameters related to the top quark are just selected  examples which underline the importance of this channel.
Also, it should be noticed that a significant fraction of the inclusive $t\bar{t}$ sample is accompanied by additional SM particles, either electroweak bosons, leptons or highly energetic jets. 
Let us focus our attention on the associated production of top-quark pairs with one hard jet (hereafter denoted $t\bar{t}j$). Besides representing a QCD background for Higgs boson searches in the Vector Boson Fusion and $t\bar{t}H$ channels, this process plays also a role in searches of physics beyond the SM (for example signals from decay chains of  SUSY particles).
The $t\bar{t}j$ production process is also important for the precision measurement of the top quark mass at the LHC  \cite{Alioli:2013mxa, Fuster:2017rev}.
We would like to stress that a precise determination of $m_t$ is crucial not only because it affects predictions of cross sections that are indispensable to  study Higgs boson properties or new signals from BSM physics, but also because the stability of the electroweak vacuum depends crucially on the actual value of this parameter \cite{Degrassi:2012ry,Alekhin:2012py}. Last but not least, the top quark mass is an ingredient for global electroweak fits which are important consistency checks of the SM \cite{Baak:2014ora}.
It is not surprising that $t\bar{t}j$ received considerable attention in the last years \cite{Dittmaier:2007wz, Dittmaier:2008uj, Melnikov:2010iu, Melnikov:2011qx, Kardos:2011qa, Alioli:2011as, Czakon:2015cla, Bevilacqua:2015qha, Bevilacqua:2016jfk}.
In this contribution we discuss the impact of off-shell effects in $t\bar{t}j$ production and show selected results from our recent work \cite{Bevilacqua:2017ipv} where we explore this channel in relation to the extraction of $m_t$ at the LHC Run II, taking the  viewpoint of a full calculation and comparing with different levels of on-shell approximations. We analyse in particular two observables, $\rho_s$ and $M_{be^+}$, which have been widely investigated both theoretically \cite{Alioli:2013mxa,Fuster:2017rev,Denner:2012yc,AlcarazMaestre:2012vp,Heinrich:2013qaa,Heinrich:2017bqp} and experimentally, see \textit{e.g.} \cite{Aad:2015waa,CMS:2016khu,Aaboud:2016igd,Sirunyan:2017idq}, with the goal of assessing their sensitivity to the top quark mass.

\section{NLO analysis of $pp \to e^+\nu_e\mu^-\bar{\nu}_\mu b \bar{b}j$}
We present here selected results for the LHC Run II, specifically NLO QCD predictions at the perturbative order $\mathcal{O}(\alpha^4\alpha_s^4)$ for the center-of-mass energy $\sqrt{s}=13$ TeV. 
Details about the SM parameters, the jet algorithm and the kinematical cuts used for the calculation are described in Ref. \cite{Bevilacqua:2017ipv}.
We employ the CT14 \cite{Dulat:2015mca}, NNPDF3 \cite{Ball:2014uwa} and MMHT2014 \cite{Harland-Lang:2014zoa} parton distribution functions, following the PDF4LHC recommendations for the LHC Run II \cite{Butterworth:2015oua}.
For the renormalization and factorization scales we consider three possibilities. The first one is the \textit{fixed} scale $\mu_R = \mu_F = \mu_0 = m_t$ while the remaining two are \textit{dynamical} scales: $\mu_R = \mu_F = \mu_0 = E_T/2$ and $\mu_R = \mu_F = \mu_0 = H_T/2$,
\begin{eqnarray}
E_T &=&\sqrt{m^2_{t}+p_T^2(t)}+\sqrt{m^2_{t}+p_T^2(\,\bar{t}\,)}\,,\\ 
H_T&=&p_T(e^+) + p_T(\mu^-) + p_T(j_{b_1}) + p_T(j_{b_2}) + p_T(j_1) +
p_T^{miss}\,,
\end{eqnarray}
where $p_T(t), \, p_T(\bar{t})$ denote the transverse momenta of the top quarks reconstructed from their decay products. Theoretical uncertainties stemming from the scale dependence of the cross section are estimated by simultaneously varying $\mu_R$ and $\mu_F$ by a factor 2 around their central value $\mu_0$. 

The goal of our analysis is to perform a systematic comparison of three distinct approaches to the calculation of $t\bar{t}j$ production in the di-lepton channel, based on different levels of approximation. The first approach, dubbed \textit{Full}, consists of a complete $\mathcal{O}(\alpha^4\alpha_s^4)$  calculation of the process $pp \to e^+\nu_e\mu^-\bar{\nu}_\mu b \bar{b}j$ where all possible contributions, \textit{i.e.} double-top , single-top, and non-resonant diagrams, are taken into account \cite{Bevilacqua:2015qha, Bevilacqua:2016jfk}. This calculation has been performed with the help of the program \texttt{HELAC-NLO} \cite{Bevilacqua:2011xh}, which comprises \texttt{HELAC-1LOOP} \cite{vanHameren:2009dr} and \texttt{HELAC-DIPOLES} \cite{Czakon:2009ss, Bevilacqua:2013iha}. The second approach, dubbed \textit{$\mbox{NWA}_{Prod}$}, considers on-shell top quarks and $W$ bosons and restricts the computation to the decay chain $pp \to t\bar{t}j \to e^+\nu_e\mu^-\bar{\nu}_\mu b \bar{b}j$ as described in Ref. \cite{Melnikov:2010iu}. This means considering on-shell $t\bar{t}j$ production at NLO while modeling spin-correlated top quark decays at LO. The third and last approach, dubbed \textit{$\mbox{NWA}$}, is a more sophisticated and complete version of narrow-width approximation which includes QCD corrections and jet radiation into top quark decays as well. This requires to take consistently into account the additional decay chain $pp \to t\bar{t} \to e^+\nu_e\mu^-\bar{\nu}_\mu b \bar{b}j$, as described in Ref. \cite{Melnikov:2011qx}.
Let us stress that this analysis is carried out at fixed order, namely effects of parton shower and hadronization are not taken into account at this stage.

The performance of different prescriptions for the renormalization and factorization scales has been extensively studied in the context of the \textit{Full} calculation. Indeed the genuine nature of $pp \to e^+\nu_e\mu^-\bar{\nu}_\mu b \bar{b}j$ as a multi-scale process suggests that a judicious choice of dynamical scales could help to capture effects from higher orders and minimize shape distortions induced by radiative corrections, thus improving the perturbative stability of our predictions. 
A comparative analysis of predictions based on different scale choices has been performed in \cite{Bevilacqua:2016jfk} considering a wide spectrum of observables. It has been shown that the fixed scale $\mu_0= m_t$ does not always ensure a stable shape when going from LO to NLO, and significant distortions have been observed particularly in the case of $p_T$ and invariant mass distributions. Also, in the fixed-scale setup the NLO error bands do not generally fit well within the LO ones as one would expect from a well-behaved perturbative expansion. Using dynamical scales, instead, the QCD corrections are positive and vary from rather small to moderate in the whole considered range.
We believe that the dynamical scale $\mu_0 = H_T/2$ performs reasonably well in accounting for the multi-scale nature of the process, at least for the kinematical setup considered in our analysis. We promote $\mu_0=H_T/2$ as the reference scale for our benchmark predictions based on the most accurate calculation, \textit{i.e.} the \textit{Full} approach. 

\begin{figure}[h!tb]
\centerline{%
\includegraphics[width=0.885\textwidth]{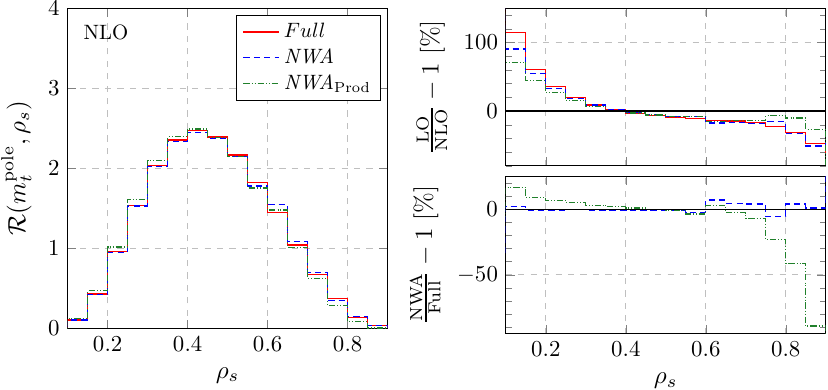}
\put(-210,103){\tiny CT14 PDF}
\put(-248,33){\scriptsize $\mu = m_t$}
}
\centerline{%
\hspace{-0.5cm} \includegraphics[width=0.95\textwidth]{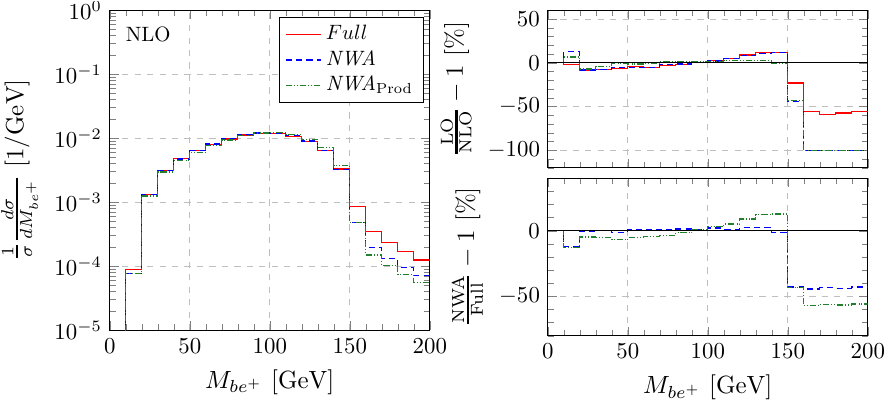}
\put(-218,106){\tiny CT14 PDF}
\put(-248,35){\scriptsize $\mu = m_t$}
}
\caption{Normalized distributions of the observables $\rho_s$ (upper plots) and $M_{be^+}$ (lower  plots) at NLO QCD. Also shown is the relative size of the QCD corrections (upper-right panels) and of the off-shell effects (lower-right panels).}
\label{Fig:Rho_Mbl_Full_vs_NWA}
\end{figure}

The overall impact of the off-shell effects related to top quark and $W$ boson decays, as comes from comparing the total NLO cross section in the \textit{Full} and \textit{NWA} approaches, is at the level of 2\%. This is fully consistent with the size of NWA effects, \textit{i.e.} $\mathcal{O}(\Gamma_t/m_t)$. It is well known, on the other hand, that this kind of effects can be dramatically enhanced in specific regions of the phase space and might play a much more relevant role at the differential level.
To assess their size on a more exclusive ground, we focus on two observables which have been widely investigated in the context of precision measurements of $m_t$ at the LHC. The first one, denoted $\mathcal{R}(m_t^{pole},\rho_s)$, is the normalized differential cross section as a function of the inverse invariant mass of the $t\bar{t}j$ system, $M_{t\bar{t}j}$  \cite{Alioli:2013mxa}:
\begin{equation}
{\cal R}(m_t^{pole},\rho_s) \equiv \frac{1}{\sigma_{t\bar{t}j}} 
\frac{d\sigma_{t\bar{t}j}}{d\rho_s} \,, 
~~~{\rm with} ~~~~
\rho_s = \frac{2m_0}{M_{t\bar{t}j}}\,,
\end{equation}
where $m_0 = 170$ GeV is a scale parameter of the order of the top quark mass. 
The second observable is the normalized differential cross section as a function of $M_{be^+}$:
\begin{equation}
{\cal R}(m_t^{pole},M_{be^+}) \equiv \frac{1}{\sigma_{t\bar{t}j}} 
\frac{d\sigma_{t\bar{t}j}}{dM_{be^+}} \,, 
~~~{\rm with} ~~~~
M_{be^+} =  \min \left\{ M_{b_1 e^+} , M_{b_2 e^+}  \right\} \,,
\end{equation}
where $b_1$ and $b_2$ denote the two $b$-jets in the final state. 
In Figure \ref{Fig:Rho_Mbl_Full_vs_NWA} we compare the NLO predictions for $\mathcal{R}(m_t^{pole},\rho_s)$ and  $\mathcal{R}(m_t^{pole},M_{be^+})$ obtained with the three different approaches. Also shown is the relative size of NLO QCD corrections and off-shell contributions on the shape of the two distributions in the full kinematical range. In the case of $\mathcal{R}(m_t^{pole},\rho_s)$, one can observe that deviations of \textit{NWA} from the \textit{Full} result are below 15\% in the most sensitive region. On the other hand, substantial differences of the order of 50\%\---100\% are visible for \textit{$\mbox{NWA}_{Prod}$} in the same region. Given that $\rho_s \approx 1$ corresponds to the threshold of $t\bar{t}$ production, which is by its own nature most sensitive to the value of $m_t$, these differences should have a considerable impact on the extraction of $m_t$ when the $\mathcal{R}(m_t^{pole},\rho_s)$ distribution is used as template for  fits.
In the case of the normalized $M_{be^+}$ distribution, a remarkably different behavior can be noticed in the regions below and above the critical value defined by $M_{be^+} = \sqrt{m_{t}^2-m_{W}^2} \approx 153$ GeV. It should be noticed that, in the \textit{$\mbox{NWA}_{Prod}$} case, this value corresponds to a kinematical endpoint for the observable at hand. When QCD radiation is included in the modeling of top quark decays, or alternatively when off-shell contributions are taken into account, the kinematical endpoint is smeared. This is the region where off-shell effects have a pretty large impact of the order of 50\% (see Figure \ref{Fig:Rho_Mbl_Full_vs_NWA}). On the contrary, they have an almost negligible size in the range below the kinematical endpoint.
%

\section{Top quark mass extraction with template methods}
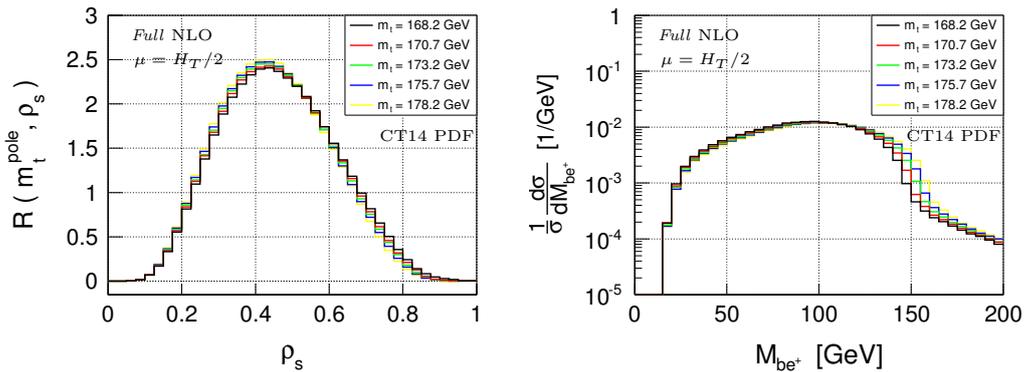
\begin{figure}[h!tb]
\centerline{%
\includegraphics[width=0.55\textwidth]{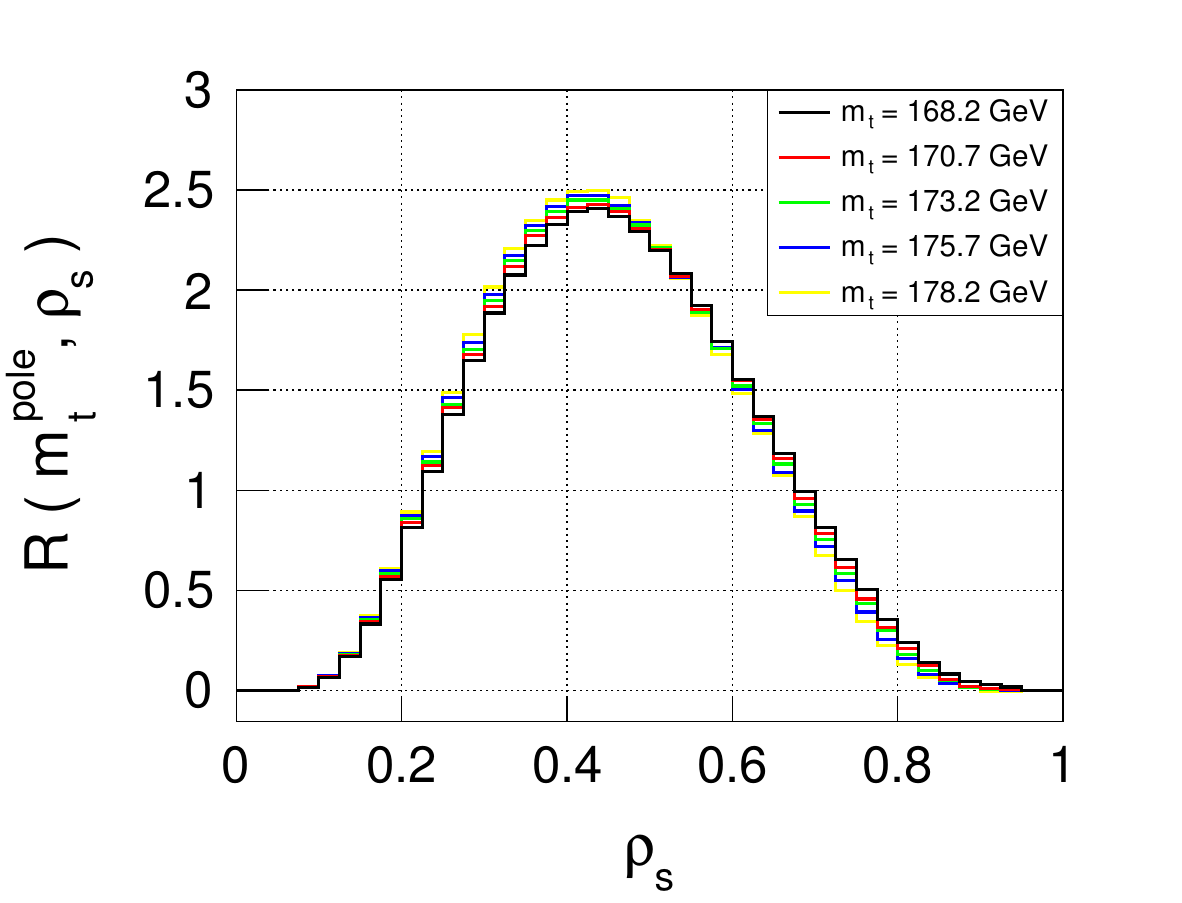}
\put(-149,126){\tiny \textit{Full} NLO}
\put(-148,117){\tiny $\mu = H_T/2$}
\put(-56,88){\tiny CT14 PDF}
\includegraphics[width=0.55\textwidth]{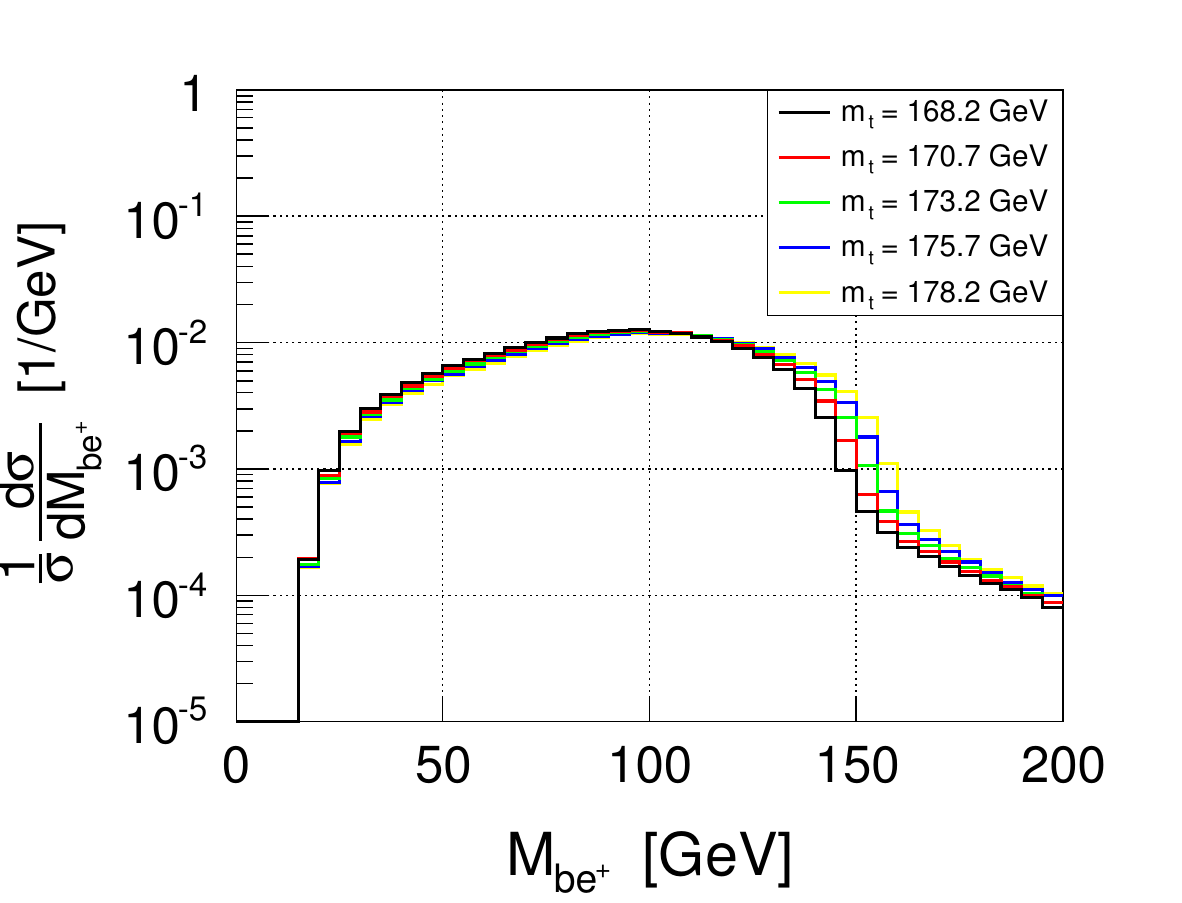}
\put(-149,126){\tiny \textit{Full} NLO}
\put(-148,117){\tiny $\mu = H_T/2$}
\put(-56,88){\tiny CT14 PDF}
}\caption{Normalized distribution of the observables $\rho_s$ (left plot) and $M_{be^+}$ (right plot) for five different values of the input $m_t$ used for the calculation.}
\label{Fig:rho_Mbl_massdep}
\end{figure}

The sensitivity of the shape of differential cross sections to the top quark mass can be exploited to extract the latter parameter from fits to data: this is the basic concept of the \textit{template method}. Figure \ref{Fig:rho_Mbl_massdep} gives an idea of  the expected variation in shape of $\mathcal{R}(m_t^{pole},\rho_s)$ and $\mathcal{R}(m_t^{pole},M_{be^+})$ for five different values of the input mass used in the calculation, ranging in steps of 2.5 GeV from $m_t=168.2$ GeV up to $m_t=178.2$ GeV.
In the case of $\mathcal{R}(m_t^{pole},\rho_s)$, as given by the most accurate predictions with $\mu_R=\mu_F=H_T/2$, a significant mass dependence can be observed in the ranges $0.25 < \rho_s < 0.45$ and $\rho_s > 0.6$.
In the case of $\mathcal{R}(m_t^{pole},M_{be^+})$ one of the most sensitive regions is the one centered around the kinematical endpoint, $140 \mbox{ GeV} < M_{be^+} <160 \mbox{ GeV}$.

To quantify the impact of the off-shell effects in this context, in the first step we generate pseudo-data for a given value of collider luminosity. These are generated according to our most accurate prediction, \textit{i.e.} the \textit{Full} calculation with $\mu_R=\mu_F=H_T/2$. Let us call this prediction "theory input" for brevity. In the second step, the pseudo-data are fitted with a template distribution, namely a prediction from either one of the three approaches that we have considered: \textit{Full}, \textit{NWA} or \textit{$\mbox{NWA}_{Prod}$}. The position of the minimum of the $\chi^2$ distribution is used to extract the numerical value of the top quark mass. To account for statistical fluctuations, the whole procedure is iterated $1000$ times. In the end a distribution of extracted masses is obtained, whose average value and spread at 68\% C.L.  define the final result of the fit in the form $m_t^{out} \pm \delta m_t^{out}$.
We refer to Ref. \cite{Bevilacqua:2017ipv} for a more detailed description of the statistical procedure.

In Table \ref{Tab:fit_Rho_Mbl} we report a comparative analysis of the results of the fit obtained for the three different approaches to the calculation, as well as for different scale choices. Results refer to the two observables described in the paper for two reference values of luminosity, $2.5 \mbox{ fb}^{-1}$ and $25 \mbox{ fb}^{-1}$, which correspond approximately to 5400 and 54000 events respectively (we have included a multiplicity factor of 4 which accounts for all combinations of charged leptons of the first two generations).
Together with the extracted mass and its uncertainty, we  monitor also the quality of the fit and the mass shift with respect to the input value $m_t^{in}$ used for the pseudo-data. 
A few comments are in order. The first thing one can notice, looking at the $\rho_s$ observable in the low-luminosity case, is an overall agreement of the pseudo-data, below 1.2$\sigma$, with any of the three approaches considered. Normally the quality of the fit would start to be questionable when an agreement worse than $2\sigma$ is found. Despite the good agreement, different mass shifts of the order of 1 GeV, 2 GeV and 3.8 GeV are observed for the \textit{Full},  \textit{NWA} and  \textit{$\mbox{NWA}_{Prod}$} cases respectively. This should be compared with the statistical uncertainty $\delta m_t^{out}$ found at low luminosity, which is of the order of 1 GeV. When the high-luminosity setup is considered, $\mathcal{L}=25 \mbox{ fb}^{-1}$, the quality of fits based on NWA results gets visibly worse. Also the template based on the \textit{Full} calculation with $\mu_0=m_t$ does not adequately describe the pseudo-data. We note that the mass shift does not change significantly with respect to our previous findings, while the statistical uncertainty $\delta m_t^{out}$ is dramatically smaller as expected. Thus, at $\mathcal{L}=25 \mbox{ fb}^{-1}$ the pseudo-data start to become sensitive to off-shell effects as well as to the scale choice. Looking now at the $M_{be^+}$ results, one can note a reduced statistical uncertainty $\delta m_t^{out}$ in comparison with $\rho_s$ for a given value of luminosity. Also lower mass shifts are observed with respect to $\rho_s$, of the order 0.2 GeV, 0.7 GeV and 0.6 GeV for \textit{Full}, \textit{NWA} and \textit{$\mbox{NWA}_{Prod}$} respectively. As observed in the case of $\rho_s$, at $\mathcal{L} = 25 \mbox{ fb}^{-1}$ the pseudo-data start to resolve off-shell and scale effects, thus templates based on the narrow-width approximation as well as the \textit{Full} case with $\mu_0 = m_t$ are clearly disfavored.
The impact of systematic uncertainties stemming from scale and PDF variations has also been estimated taking the \textit{Full} case as a benchmark.  Concerning the $\rho_s$ observable, scale uncertainties are of the order of 2 GeV for the $\mu_0=m_t$ setup, while they are reduced to 0.6 \-- 1.2 GeV when dynamical scales are used. In the case of $M_{be^+}$ they are smaller, of the order of 1 GeV for the fixed scale and only 0.05 GeV for dynamical scales. On the other hand, PDF uncertainties are at the level of 0.4 \-- 0.7 GeV for $\rho_s$ and 0.02 \-- 0.03 GeV for $M_{be^+}$ independently on the scale choice, therefore they are well below the dominant scale systematics.

\begin{table}[t!]
\begin{center}
\begin{tabular}{ccccc}
\hline\hline
&&&&\\
{Theory, NLO  QCD}& $m^{out}_t \pm \delta m^{out}_t$
& Average & Probability &
$m_t^{in} -m_t^{out}$ \\ {CT14 PDF} & [GeV] &
$\chi^2/{\rm d.o.f.}$
& {\it p-value} & [GeV] \\
&&&&\\
\hline\hline
&&&&\\
$\rho_s$  &$\mathcal{L} = 2.5 \mbox{ fb}^{-1}$&&&\\ \vspace{-0.3cm}
&&&&\\
\hline\hline
{\it Full}, $\mu_0=H_T/2$& 173.05 $\pm$ 1.31 & 0.99 
&0.42 (0.8$\sigma$)& $+0.15$\\
{\it Full}, $\mu_0=E_T/2$ &  172.19 $\pm$ 1.34 & 1.05
 & 0.39 (0.9$\sigma$) &$+1.01$\\
{\it Full}, $\mu_0=m_t$ & 173.86 $\pm$ 1.39   & 1.42 
& 0.21 (1.2$\sigma$) & $-0.66$\\
\hline\hline 
{\it NWA},  $\mu_0=m_t$ & 175.22 $\pm$ 1.15 &
1.38 & 0.23 (1.2$\sigma$)&  $-2.02$\\
{\it NWA}${}_{Prod.}$, $\mu_0=m_t$ & 169.39  $\pm$ 1.46  &
1.12 & 0.35 (0.9$\sigma$) &  $+3.81$\\
\hline\hline
&&&&\\
$\rho_s$ &$\mathcal{L} = 25 \mbox{ fb}^{-1}$&&&\\ \vspace{-0.3cm}
&&&&\\
\hline\hline
{\it Full}, $\mu_0=H_T/2$& 173.06 $\pm$ 0.44 & 0.97 
&0.44 (0.8$\sigma$)& $+0.14$\\
{\it Full}, $\mu_0=E_T/2$ &  172.36 $\pm$ 0.44 & 1.38
 & 0.23 (1.2$\sigma$) &$+0.84$\\
{\it Full}, $\mu_0=m_t$ & 173.84 $\pm$ 0.42   & 5.12 
& 1 $\cdot 10^{-4}$ (3.9$\sigma$) & $-0.64$\\
\hline\hline 
{\it NWA},  $\mu_0=m_t$ & 175.23 $\pm$ 0.37 &
5.28 & 7 $\cdot 10^{-5}$ (4.0$\sigma$)&  $-2.03$\\
{\it NWA}${}_{Prod.}$, $\mu_0=m_t$ & 169.43  $\pm$ 0.50  &
2.61 & 0.02 (2.3$\sigma$) &  $+3.77$\\
\hline\hline 
&&&&\\
$M_{be^+}$ & $\mathcal{L} = 2.5 \mbox{ fb}^{-1}$  &&&\\ \vspace{-0.3cm}
&&&&\\
\hline\hline
{\it Full}, $\mu_0=H_T/2$&173.09 $\pm$ 0.48 & 1.05 
& 0.38 (0.9$\sigma$) & $+$0.11\\
{\it Full}, $\mu_0=E_T/2$& 173.01 $\pm$ 0.50 & 1.06
& 0.37 (0.9$\sigma$) & $+$0.19\\
{\it Full}, $\mu_0=m_t$& 173.07 $\pm$ 0.49 & 1.22 
& 0.18 (1.3$\sigma$) &$+$0.13\\
\hline\hline
{\it NWA},  $\mu_0=m_t$&173.90 $\pm$ 0.50& 1.11
& 0.30 (1.0$\sigma$) &$-$0.70 \\
{\it NWA}$_{\rm Prod.}$, $\mu_0=m_t$& 172.56 $\pm$ 0.54& 1.64
& 0.01 (2.6$\sigma$) &$+$0.64 \\
\hline\hline
&&&&\\
$M_{be^+}$  & $\mathcal{L} = 25 \mbox{ fb}^{-1}$  &&&\\ \vspace{-0.3cm}
&&&&\\
\hline\hline
{\it Full}, $\mu_0=H_T/2$& 173.18 $\pm$ 0.15 &
1.02  & 0.42 (0.8$\sigma$) & $+$0.02 \\
{\it Full}, $\mu_0=E_T/2$& 173.23 $\pm$ 0.15 & 1.03
& 0.41  (0.8$\sigma$)&$-$0.03\\
{\it Full}, $\mu_0=m_t$& 173.22 $\pm$ 0.16 & 1.78
& 0.005  (2.8$\sigma$) &$-$0.02\\
\hline\hline
{\it NWA},  $\mu_0=m_t$& 173.98  $\pm$ 0.16 & 2.56 
& 5 $\cdot 10^{-6}$ (4.6$\sigma$) & $-$0.78\\
{\it NWA}$_{\rm Prod.}$, $\mu_0=m_t$&  172.62 $\pm$ 0.17 & 8.23
& 0 ($\gg 5\sigma$) & $+$0.58\\
\hline\hline
\end{tabular}
\end{center}
\caption{\it Top quark mass fits obtained using the normalized $\rho_s$ and $M_{be^+}$ distributions as templates. Results are shown for the two reference luminosities of 2.5 $fb^{-1}$ and 25 $fb^{-1}$. From left to right: mean value of the top quark mass ($m_t^{out}$) obtained from 1000 pseudo-data sets together with its 68$\,\%$ C.L. statistical error ($\delta m_t^{out}$); average minimum $\chi^2/d.o.f$; $p$-value with the corresponding number of standard deviations; top quark mass shift ($m_t^{in} -m_t^{out}$). For the $\rho_s$ distribution, the histogram binning of Ref.\cite{Aad:2015waa} is considered.}
\label{Tab:fit_Rho_Mbl}
\end{table}
%

\section{Conclusions}
In this paper we have studied the normalized distributions of two observables of interest in the study of $t\bar{t}j$ production with leptonic decays at the LHC, namely $\rho_s$ and $M_{be^+}$. Through a systematic comparison of the full fixed-order calculation with results based on NWA, we have found that the off-shell effects play an important role in kinematical regions which are relevant for the extraction of $m_t$. Overall, the $M_{be^+}$ observable shows the best performance in terms of the  statistical uncertainty on the extracted top quark mass as well as the systematics related to scale and PDF variations. Our fixed-order analysis indicates that off-shell effects have an impact on the fits particularly in the high-luminosity case considered,  $\mathcal{L}=25 \mbox{ fb}^{-1}$. Templates based on the narrow-width approximation induce visible mass shifts which are relevant also at lower luminosities. 
The results presented in this paper are part of a wider phenomenological study aimed at exploring the relevance of the off-shell effects on a more extensive set of kinematical observables. Although not explicitly shown here, we have found other observables, such as $M_{t\bar{t}}$ and $H_T$, which exhibit competitive performances for the purpose of extracting $m_t$  from $t\bar{t}j$ \cite{Bevilacqua:2017ipv}.

\medskip \medskip
The research of G.~B. was supported by grant K 125105 of the National Research, Development and Innovation Office in Hungary. The work of M.~W. and H.~B.~H. was supported in part by the German Research Foundation (DFG) under Grant no. WO 1900/2 \-- \textit{Top Quarks under the LHCs Magnifying Glass: From Process Modeling to Parameter Extraction}.
The work of H.~B.~H. was also supported by a Rutherford Grant ST/M004104/1.


\begin{thebibliography}{99}

\bibitem{Czakon:2013goa}
  M.~Czakon, P.~Fiedler and A.~Mitov,
  Phys.\ Rev.\ Lett.\  {\bf 110} (2013) 252004.

\bibitem{Czakon:2016olj}
  M.~Czakon, N.~P.~Hartland, A.~Mitov, E.~R.~Nocera and J.~Rojo,
  JHEP {\bf 1704} (2017) 044.

\bibitem{Alioli:2013mxa}
  S.~Alioli \textit{et al.},
  Eur.\ Phys.\ J.\ C {\bf 73} (2013) 2438.

\bibitem{Fuster:2017rev}
  J.~Fuster, A.~Irles, D.~Melini, P.~Uwer and M.~Vos,
  arXiv:1704.00540 [hep-ph].

\bibitem{Degrassi:2012ry}
  G.~Degrassi \textit{et al.},
  JHEP {\bf 1208} (2012) 098.

\bibitem{Alekhin:2012py}
  S.~Alekhin, A.~Djouadi and S.~Moch,
  Phys.\ Lett.\ B {\bf 716} (2012) 214.

\bibitem{Baak:2014ora}
  M.~Baak {\it et al.} [Gfitter Group],
  Eur.\ Phys.\ J.\ C {\bf 74} (2014) 3046.

\bibitem{Dittmaier:2007wz}
  S.~Dittmaier, P.~Uwer and S.~Weinzierl,
  Phys.\ Rev.\ Lett.\  {\bf 98} (2007) 262002.

\bibitem{Dittmaier:2008uj}
  S.~Dittmaier, P.~Uwer and S.~Weinzierl,
  Eur.\ Phys.\ J.\ C {\bf 59} (2009) 625.

\bibitem{Melnikov:2010iu}
  K.~Melnikov and M.~Schulze,
  Nucl.\ Phys.\ B {\bf 840} (2010) 129.

\bibitem{Melnikov:2011qx}
  K.~Melnikov, A.~Scharf and M.~Schulze,
  Phys.\ Rev.\ D {\bf 85} (2012) 054002.

\bibitem{Kardos:2011qa}
  A.~Kardos, C.~Papadopoulos and Z.~Trocsanyi,
  Phys.\ Lett.\ B {\bf 705} (2011) 76.

\bibitem{Alioli:2011as}
  S.~Alioli, S.~O.~Moch and P.~Uwer,
  JHEP {\bf 1201} (2012) 137.

\bibitem{Czakon:2015cla}
  M.~Czakon, H.~B.~Hartanto, M.~Kraus and M.~Worek,
  JHEP {\bf 1506} (2015) 033.

\bibitem{Bevilacqua:2015qha}
  G.~Bevilacqua, H.~B.~Hartanto, M.~Kraus and M.~Worek,
  Phys.\ Rev.\ Lett.\  {\bf 116} (2016) no.5,  052003.

\bibitem{Bevilacqua:2016jfk}
  G.~Bevilacqua, H.~B.~Hartanto, M.~Kraus and M.~Worek,
  JHEP {\bf 1611} (2016) 098.

\bibitem{Bevilacqua:2017ipv}
  G.~Bevilacqua, H.~B.~Hartanto, M.~Kraus, M.~Schulze and M.~Worek,
  arXiv:1710.07515 [hep-ph].

\bibitem{Denner:2012yc}
  A.~Denner, S.~Dittmaier, S.~Kallweit and S.~Pozzorini,
  JHEP {\bf 1210} (2012) 110.

\bibitem{AlcarazMaestre:2012vp}
  J.~Alcaraz Maestre {\it et al.} [SM and NLO MULTILEG Working Group and SM MC Working Group],
  arXiv:1203.6803 [hep-ph].

\bibitem{Heinrich:2013qaa}
  G.~Heinrich \textit{et al.},
  JHEP {\bf 1406} (2014) 158.

\bibitem{Heinrich:2017bqp}
  G.~Heinrich \textit{et al.},
  arXiv:1709.08615 [hep-ph].

\bibitem{Aad:2015waa}
  G.~Aad {\it et al.} [ATLAS Collaboration],
  JHEP {\bf 1510} (2015) 121.
  
\bibitem{CMS:2016khu}
  CMS Collaboration [CMS Collaboration],
  CMS-PAS-TOP-13-006.

\bibitem{Aaboud:2016igd}
 M.~Aaboud {\it et al.} [ATLAS Collaboration],
 Phys.\ Lett.\ B {\bf 761} (2016) 350. 

\bibitem{Sirunyan:2017idq}
  A.~M.~Sirunyan {\it et al.} [CMS Collaboration],
   arXiv:1704.06142 [hep-ex].

\bibitem{Dulat:2015mca}
  S.~Dulat {\it et al.},
  Phys.\ Rev.\ D {\bf 93} (2016) no.3,  033006.

\bibitem{Ball:2014uwa}
  R.~D.~Ball {\it et al.} [NNPDF Collaboration],
  JHEP {\bf 1504} (2015) 040.

\bibitem{Harland-Lang:2014zoa}
  L.~A.~Harland-Lang \textit{et al.},
  Eur.\ Phys.\ J.\ C {\bf 75} (2015) 5,  204.
  
\bibitem{Butterworth:2015oua}
  J.~Butterworth {\it et al.},
  J.\ Phys.\ G {\bf 43} (2016) 023001.

\bibitem{Bevilacqua:2011xh}
  G.~Bevilacqua \textit{et al.},
  Comput.\ Phys.\ Commun.\  {\bf 184} (2013) 986.

\bibitem{vanHameren:2009dr}
  A.~van Hameren, C.~G.~Papadopoulos and R.~Pittau,
  JHEP {\bf 0909} (2009) 106.

\bibitem{Czakon:2009ss}
  M.~Czakon, C.~G.~Papadopoulos and M.~Worek,
  JHEP {\bf 0908} (2009) 085.

\bibitem{Bevilacqua:2013iha}
  G.~Bevilacqua, M.~Czakon, M.~Kubocz and M.~Worek,
  JHEP {\bf 1310} (2013) 204.

\end{thebibliography}
\end{document}